\documentclass[conference]{IEEEtran}
\IEEEoverridecommandlockouts
\usepackage{cite}
\usepackage{amsmath,amssymb,amsfonts}
\usepackage{algorithmic}
\usepackage{graphicx}
\usepackage{textcomp}
\usepackage{xcolor}
\usepackage{booktabs}
\usepackage{hyperref}
\usepackage{algorithm}
\def\BibTeX{{\rm B\kern-.05em{\sc i\kern-.025em b}\kern-.08em
    T\kern-.1667em\lower.7ex\hbox{E}\kern-.125emX}}
\begin{document}

\title{OpenGCRAM: An Open-Source Gain Cell Compiler Enabling Design-Space Exploration for AI Workloads
}

\author{%
  \IEEEauthorblockN{%
    Xinxin Wang\IEEEauthorrefmark{1},\,
    Lixian Yan\IEEEauthorrefmark{1},\,
    Shuhan Liu\IEEEauthorrefmark{1},\,
    Luke Upton\IEEEauthorrefmark{1},\,
    Zhuoqi Cai\IEEEauthorrefmark{1},\,
    Yiming Tan\IEEEauthorrefmark{1},\,
    Shengman Li\IEEEauthorrefmark{1},\\
    Koustav Jana\IEEEauthorrefmark{1},\,
    Peijing Li\IEEEauthorrefmark{1},\,
    Jesse Cirimelli–Low\IEEEauthorrefmark{2},\,
    Thierry Tambe\IEEEauthorrefmark{1},\,
    Matthew Guthaus\IEEEauthorrefmark{2},\,
    H.-S.\ Philip Wong\IEEEauthorrefmark{1}
  }
  \IEEEauthorblockA{%
    \IEEEauthorrefmark{1}Stanford University
  \IEEEauthorblockA{%
    \IEEEauthorrefmark{2}University of California, Santa Cruz\\
    \texttt{\{xxwang1,hspwong\}@stanford.edu}
  }
  }
}


\maketitle
\section{absrtact}
\begin{abstract}
Gain Cell memory (GCRAM) offers higher density and lower power than SRAM, making it a promising candidate for on-chip memory in domain-specific accelerators. To support workloads with varying traffic and
lifetime metrics, GCRAM also offers high bandwidth, ultra low leakage power and a wide range of retention times, which can be adjusted through transistor design (like threshold voltage and channel material) and on-the-fly by changing the operating voltage. However, designing and optimizing GCRAM sub-systems can be time-consuming. In this paper, we present OpenGCRAM, an open-source GCRAM compiler capable of generating GCRAM bank circuit designs and DRC- and LVS-clean layouts for commercially available foundry CMOS, while also providing area, delay, and power simulations based on user-specified configurations (e.g., word size and number of words). OpenGCRAM enables fast, accurate, customizable, and optimized GCRAM block generation, reduces design time, ensure process compliance, and delivers performance-tailored memory blocks that meet diverse application requirements.
\end{abstract}

\begin{IEEEkeywords}
Gain Cell memory (GCRAM), SRAM, memory compiler, design space exploration, AI workloads
\end{IEEEkeywords}

\section{Introduction}

The energy and delay consumed by off-chip memory (DRAM) and memory-to-logic chip data movement has become the bottleneck (known as the memory wall~\cite{gholami_ai_2024}) for modern computing systems, especially for neural network accelerators and abundant-data computing. Providing high-capacity, high-bandwidth on-chip memory is a solution. GCRAM with 2 transistors (2T) per cell offers higher density and thus larger capacity than 6T SRAM for the same chip area~\cite{complementary_gc1, complementary_gc3, liu2023gain}. This density benefit is scalable to FinFET technology nodes \cite{16nmGC}. It also consumes less power and has bandwidth comparable to SRAM \cite{wwl_ls}, making GCRAM a good complement for on-chip memory. 

GCRAM has many variants. 3T GCRAM adds one more transistor at the read path to enhance sense margin~\cite{3tgc}, and 4T GCRAM adds feedback transistor to enhance retention, but at the cost of extra area~\cite{4tgc}. To keep the high-density benefit of 2T GCRAM, we can also leverage transistor engineering (e.g. threshold voltage) or material choice (e.g. oxide semiconductor (OS) for the channel material) to provide tunability. OS channel material has large band gap that enables ultra-low off-state leakage current (\textless1e-18 A/$\mathrm{\mu m}$), leading to longer retention, and more importantly, a large design trade-off space \cite{complementary_gc3, liu2023gain, ye2020double}. The retention can range from microseconds to hours, thus providing a retention-speed trade-off to tailor the design for specific AI tasks and accelerator architectural designs. In addition to the cell-level variants, peripheral circuit design (e.g. with or without level shifter) and memory bank architecture design (e.g. array size) are also critical for memory optimization, especially for large memory capacity. With multiple design choices from material to circuit, GCRAM macros offer extensive tunability in speed, retention and area to serve different applications. However, full-stack optimization with the vast design space is challenging with exponentially increased design complexity. Manually designing the circuit blocks in analog environments is time consuming and error-prone.   

Inspired by the OpenRAM SRAM compiler~\cite{OpenRAM}, this work introduces OpenGCRAM, a GCRAM compiler that can generate GCRAM macro based on user-specified configurations. The main contributions of this paper are listed below:


\begin{itemize}
    \item We formulate a standardized methodology for porting the OpenRAM SRAM compiler to advanced technology nodes and incorporate various types of memory technologies beyond SRAM.
    \item We successfully port OpenGCRAM to TSMC 40 nm technology, resolving all DRC and LVS errors in the layout generation process. Porting to other advanced technology nodes can follow a similar methodology.
    \item We develop OpenGCRAM for GCRAM compilation, enabling the generation of SPICE netlists, layout GDS files, and precise HSPICE simulations for GCRAM banks. This facilitates fast and accurate design optimization tailored to diverse application requirements. The generated GDS files are ready for tape out with TSMC.
    \item We integrate OS-OS GCRAM circuits and layout generation into OpenGCRAM by including a compact device model and adding custom layer definitions and design rules.
    \item We make OpenGCRAM completely user-modifiable. Our code is available at \url{https://github.com/xxwang1/OpenGCRAM.git}~\footnote{Except for the TSMC 40 nm tech script and proprietary custom cells which are protected by TSMC non-disclosure agreements.}.
\end{itemize}

\section{Related Work}
\subsection{OpenRAM for Compiling SRAM Designs}
\label{sec:openram}
OpenRAM is an open-source memory compiler that supports the front-end and back-end design flows, including SPICE netlist and layout generation, characterization and verification \cite{OpenRAM}. Although it was originally developed for SRAM design, it is flexible and portable for other memory types and technologies. In addition, OpenRAM supports compiling designs at different PVT conditions, which provides insights for post-silicon results. 

In OpenRAM, the architectural building blocks of a memory macro consist of the bitcell array, address decoder, wordline (WL) driver, column multiplexer, bitline (BL) precharge, sense amplifier, write driver and control logic~\cite{OpenRAM}. With user-defined specifications like word size, number of words, number of banks, number and type of ports, and technology parameters, OpenRAM compiles a logical SPICE/Verilog model for functional simulation tests and characterizes the design to extract timing, area, and power metrics. Modular result files are retained along with the compilation. After the design passes a set of regression tests, DRC checks, and LVS checks, OpenRAM stores all output files in a folder containing the high-level layout, the Liberty files, a full netlist, and a trimmed netlist for shorter simulation time.

OpenRAM users can import customized memory cells or circuit blocks. Examples of potential custom imported blocks include a customized memory cell with different specifications and a well-designed sense amplifier for low-power operation. Besides the design adaptability, OpenRAM is developed with programmability in mind, from parametrized transistors and logic gates to delay chains and so on. Typically, a customized memory cell may require resizing the driving circuits. OpenRAM implements the logical effort and path delay calculation to resize the driving gates \cite{OpenRAM}. At present, this flexibility and portability is mainly applicable to 6T, 8T, or 10T SRAMs. Since a GCRAM has separate read and write ports, its operation and control are very different from the SRAM. Thus, changes from the cell level to the architectural level are necessary to adapt this versatile compiler for GCRAM. In Section~\ref{sec:methodology}, we will discuss the key modifications that extend OpenRAM to OpenGCRAM, facilitating a full-stack scalable GCRAM design.
\subsection{RAAAM's GCRAM Technology}
According to RAAAM Memory Technologies, their GCRAM technology is a cost-effective on-chip memory solution that delivers up to 50\% silicon area reduction and 10x reduced power consumption compared to 6T SRAM~\cite{GCRAM}. Their GCRAM macro is fully compatible with standard CMOS fabrication flows, and its design leverages a 3T GCRAM for a two-port memory, reducing transistor count and thus array area by half, compared to the 6T single-port SRAM bit-cell. The GCRAM macro has decoupled read and write paths, enabling native two-port operation and significantly increasing memory bandwidth. Furthermore, it achieves voltage scaling down to 450 mV in advanced nodes such as 16nm FinFET for substantial power savings~\cite{GCRAM}. This technology has already been validated on silicon in process nodes ranging from 16 nm – 180 nm and successfully evaluated in 5 nm FinFET technology.

Despite GCRAM's remarkable compatibility, area savings, and energy efficiency, it is patented and commercialized. The research community does not have open access to their GCRAM IP. This limits innovation and exploration of GCRAM and its ubiquitous deployment.

\subsection{GEMTOO for Design Sapce Exploration}
GEMTOO is a high-level modeling tool for design space exploration of GCRAM embedded DRAM (eDRAM). With technology parameters, circuit specification, and memory architecture, GEMTOO estimates the timing, memory availability, bandwidth, and area of a GCRAM macro for eDRAM~\cite{GEMTOO}. However, an important performance metric, namely, power evaluation, is missing in GEMTOO.  Since GEMTOO focuses on fast exploration in the design space, the chip area is estimated from the cell area and the circuit block area. Even though floorplanning is considered, it overlooks the area margin reserved to pass the DRC checks. Likewise, its timing estimation considers the worst-case scenario for read and write, and the total delay is calculated from the RC loading, such as wire resistance and wire capacitance, and characterized gate delays. Since GEMTOO avoids netlisting and SPICE-level simulations, it does not account for delays caused by layout parasitics and routing. As a result, memory metrics outputted by GEMTOO have up to 15\% deviation from the post-layout simulations~\cite{GEMTOO}. While GEMTOO excels in fast architectural prototyping, it is not meant to assist the entire design flow of GCRAM through the tapeout stage, as the back-end implementations are challenging and time consuming. The research community needs a compiler beyond analytical modeling. 
%




\section{Methodology}
\label{sec:methodology}

As memory technologies advance rapidly, memory compilers with enhanced features are essential to support the design of memory banks optimized for a variety of application requirements. In this section, we introduce a methodology for extending OpenRAM to accommodate new technology nodes and memory circuit designs.
\begin{figure}[ht]
    \centering
    \includegraphics[width=0.4\textwidth]{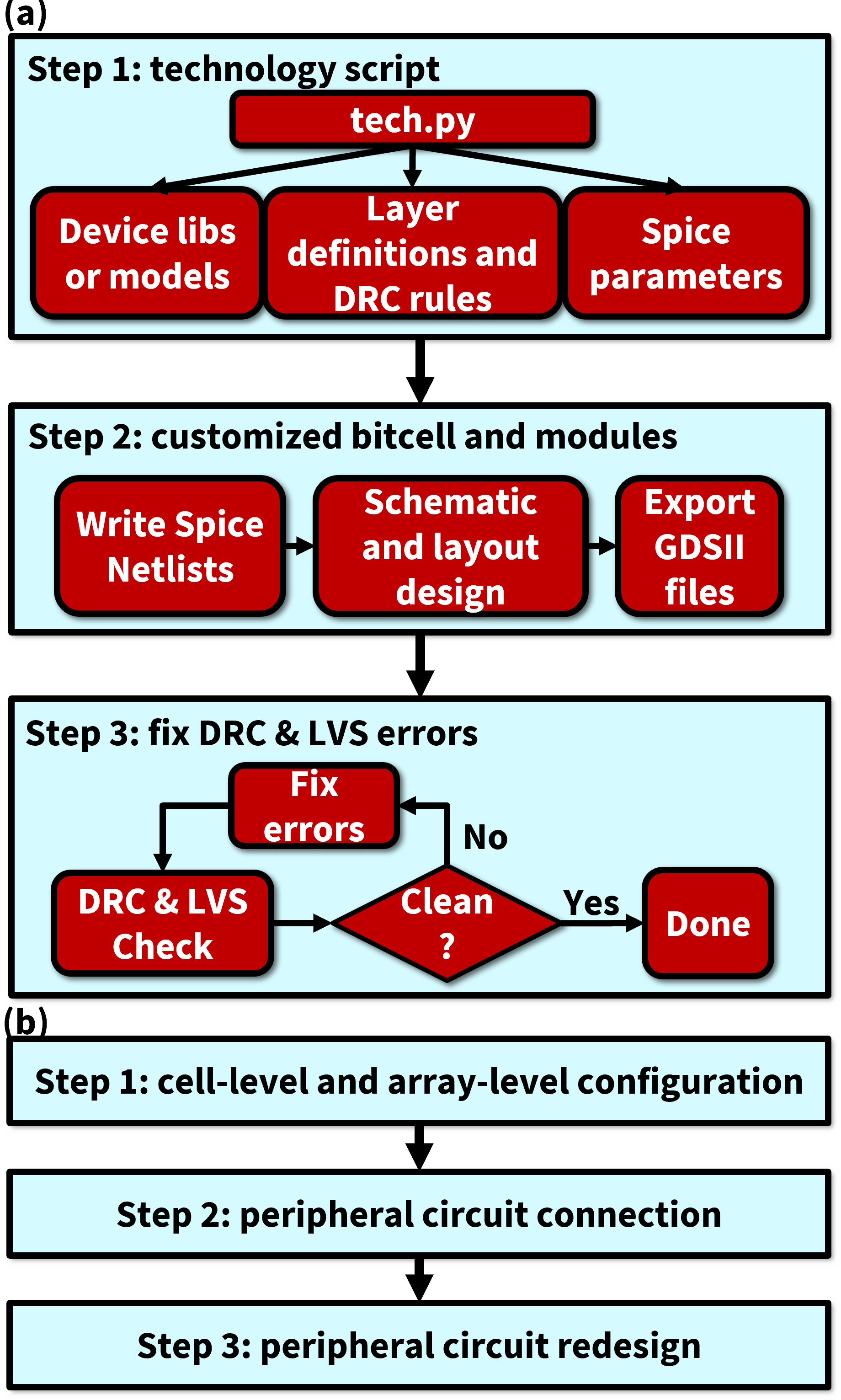} 
    \caption{Flow chart of (a) porting the compiler to new technology nodes and (b) adding support for new memory technologies.}
    \label{fig:flow_chart}
    \vspace{-10pt}
\end{figure}
\subsection{Porting to New Technologies}
\label{sec:porting}
OpenRAM currently supports SRAM and ROM compilation across several open-source process design kits (PDKs). However, modern chip designs typically require compatibility with advanced commercial PDKs, which can be achieved by following the key steps described in Fig.~\ref{fig:flow_chart}(a). First, a Python script must be added to include essential configuration details. For circuit SPICE netlist generation, the script requires paths to the transistor libraries or models; for layout generation, it needs layer definitions and basic design rules; and for analytical delay analysis, it must specify basic parameters such as wire resistance and capacitance. 

Second, the SPICE netlists and layouts for core modules, such as the bitcell, write driver, sense amplifier, and D-flip flop (DFF), should be manually created and integrated into the compiler, while other modules can be generated automatically by the compiler. If additional custom modules are required, developers can easily integrate these through the method outlined in Section~\ref{sec:new_mem}. Note that the manual design of custom modules is only a part of the compiler development process. Once the compiler is set up, the generation and simulation of memory banks will be fully automated, requiring no further manual design. 

Third, developers should set the configurations and technology process information for a memory macro and run simulations and DRC and LVS checks using the compiler to verify that the generated circuit meets design rules. Different PDKs may impose varying requirements for layer widths, spacing, area, extensions, or enclosures. Many of these rules can be automatically satisfied by incorporating basic design rules into the Python script, following existing examples in OpenRAM. However, more advanced technologies may introduce additional design rules that developers will need to define and integrate manually. If the DRC and LVS check reports any errors, they are most likely due to logic gate generation or the placement and routing of different modules. To achieve a clean DRC and LVS, developers may need to adjust the layout generation functions in the Python code.
\subsection{Adding Support for New Memory Types}
\label{sec:new_mem}
Different memory technologies have different port configurations and require different peripheral circuits. For example, an SRAM cell can have either a single port or dual ports, both utilizing differential bitlines (BL and $\mathrm{BL_{b}}$)
for readout. For GCRAM, there are separate write wordline (WWL), write bitline (WBL), read wordline (RWL) and read bitline (RBL). A one-transistor-one-capacitor (1T1C) DRAM cell consists of a single BL and WL. One-transistor-one-resistor (1T1R) RRAM includes an additional source line (SL) alongside the BL and WL. To extend the capability of OpenRAM with a new memory technology, the developers can follow the procedures in Fig.~\ref{fig:flow_chart}(b). First, they need to manually create the layout and netlist of the bitcell, then add its definition to the custom cell lists. Following this, the bitcell python script should be modified to configure its read and write ports, and the bitcell array script should be updated to generate the bitcell array by connecting the appropriate WLs, BLs and/or SLs. 

In addition to modifying the bitcell array, the connection between the array and peripheral circuits should also be adjusted if the bitcell configuration differs from that of SRAM. For instance, since GCRAM has only one BL per port, the $\mathrm{BL_{b}}$ connection should be removed. Similarly, for a 1T1R RRAM, the source line (SL) must  be connected to the peripheral circuits.

The peripheral circuits may need to be redesigned for new memory technologies, as their read and/or write operations can differ from OpenRAM. For example, the WL and BL voltage polarity may be different. The single-ended read out need additional reference voltage or current signals. SLs should also be driven by peripheral circuits. To implement this functionality, a new Python script should be added to the compiler to generate the necessary transistors, connections, pins, and layouts for this module. At a higher level, these new modules should be integrated into either the port address circuit (for driving the WLs) or the port data circuit (for driving the BLs). This integration can be accomplished by modifying the Python functions that handle module and pin list creation, internal pin connections, and the placement and routing of layouts. 

\section{Results}
\subsection{ GCRAM Bank Architecture}
\begin{figure}[ht]
    \centering
    \includegraphics[width=0.475\textwidth]{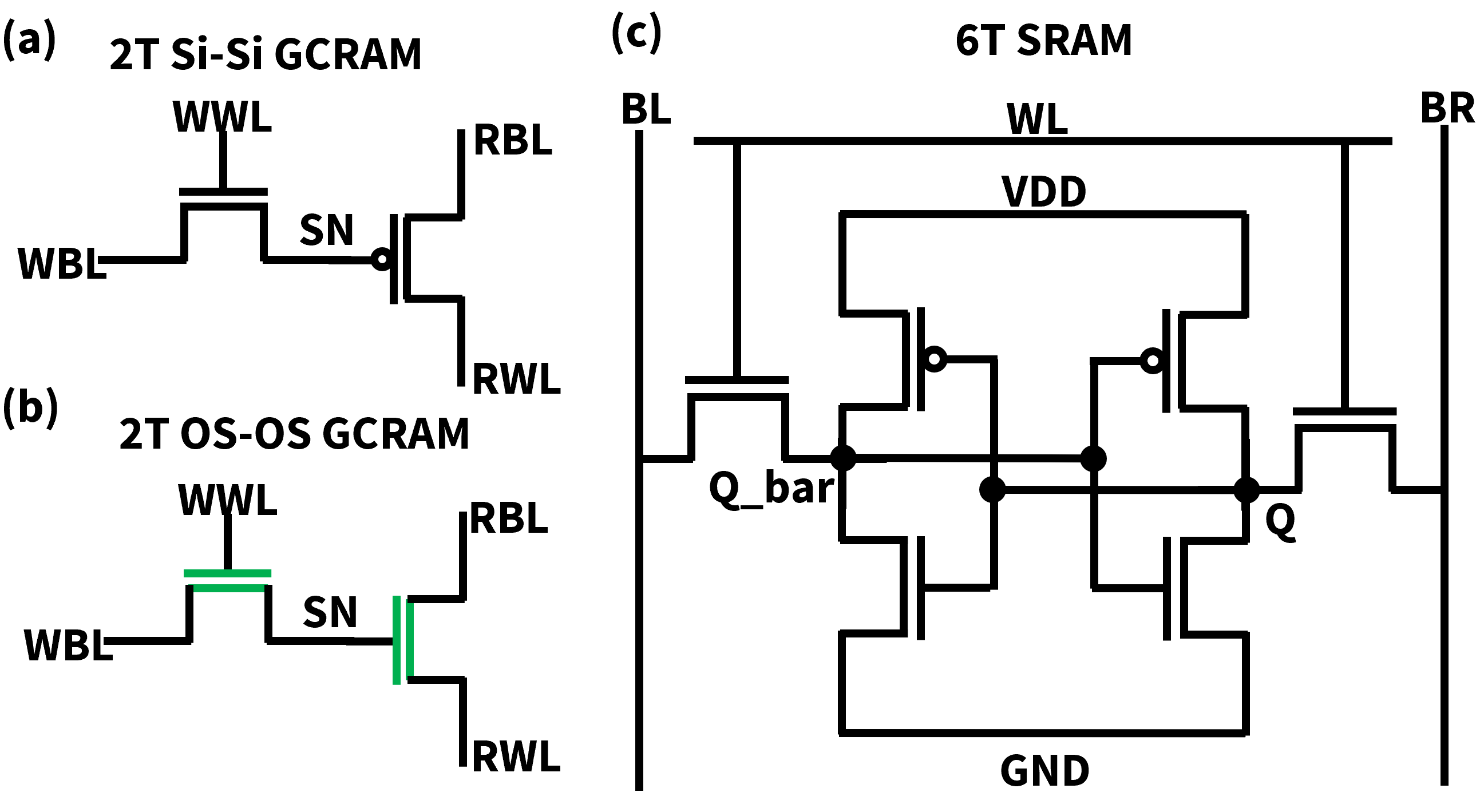} 
    \caption{The schematics of (a) 2T Si-Si GCRAM, (b) 2T OS-OS GCRAM (the green transistors denote OS transistors) and (c) 6T-SRAM which are all implemented in our compiler. }
    \label{fig:gc_sram_schematic}
    \vspace{-6pt}
\end{figure}

\begin{figure}[ht]
    \centering
    \includegraphics[width=0.475\textwidth]{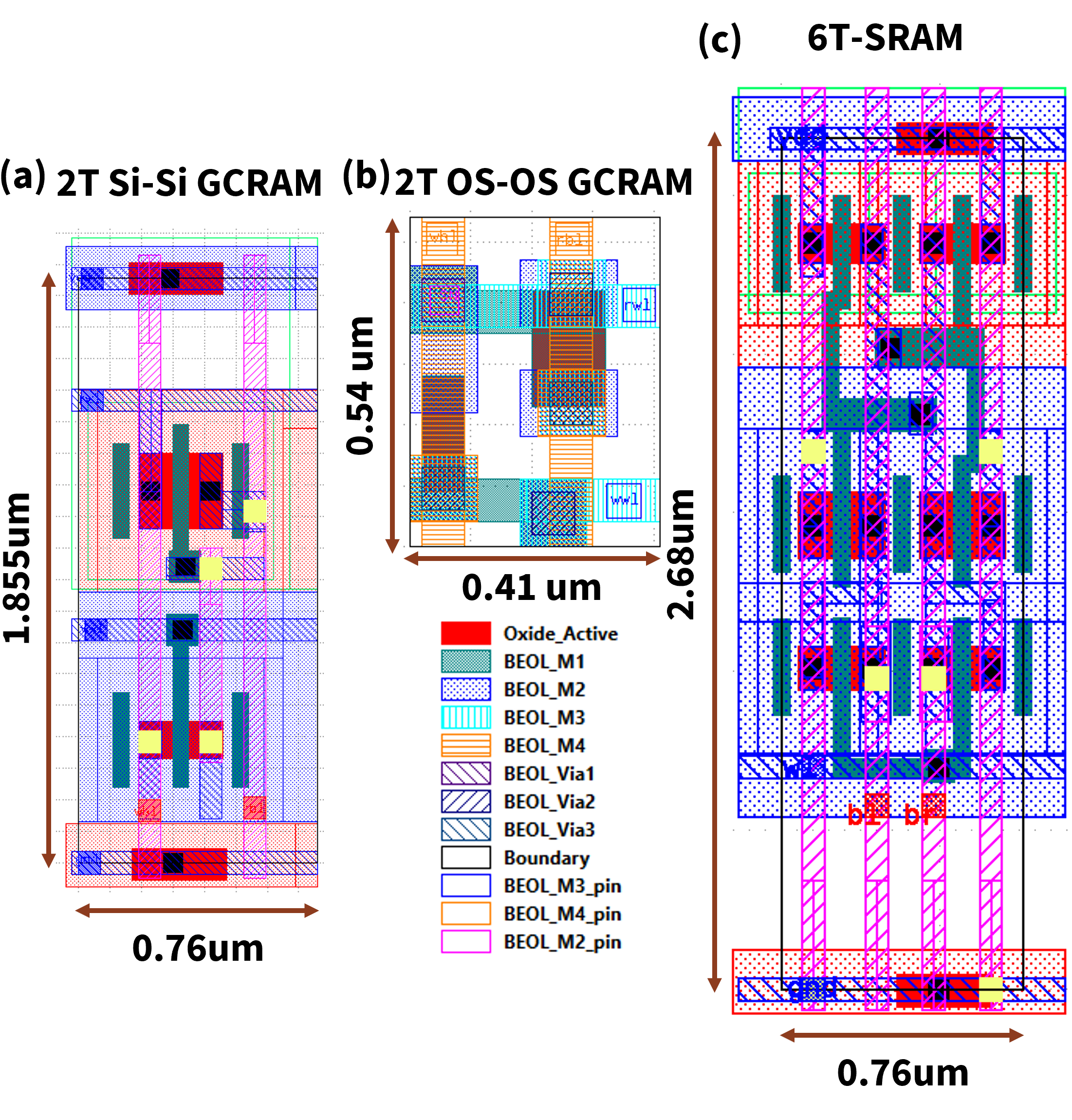} 
    \caption{The layouts of (a) 2T Si-Si GCRAM, (b) 2T OS-OS GCRAM and (c) 6T-SRAM. The Si-Si GCRAM and SRAM layouts are designed using the logic design rule. The OS-OS GCRAM is fabricated between tight-pitched metal layers, and thus the layout meets the basic FEOL design rules regarding width, space, enclosure and extension. The Si-Si GCRAM and OS-OS GCRAM occupy 69\% and 11\% of the area of a 6T SRAM cell, respectively.}
    \label{fig:gc_sram_layout}
    \vspace{-6pt}
\end{figure}

Built upon OpenRAM, OpenGCRAM successfully integrates support for TSMC 40 nm technology by following the procedures discussed in Section~\ref{sec:porting} and illustrated in Fig.~\ref{fig:flow_chart}. This process involved creating a technology file that includes device models, layer properties, basic design rules, and SPICE parameters, editing the layout generation functions and fix DRC and LVS errors. Fig.~\ref{fig:gc_sram_schematic}(a) and (b) show schematics of a 2T GCRAM with silicon transistors (2T Si-Si GCRAM) and oxide semiconductor transistors (2T OS-OS GCRAM). The 6T SRAM schematic is shown in Fig.~\ref{fig:gc_sram_schematic}(c). The 2T Si-Si GCRAM layout, shown in Fig.~\ref{fig:gc_sram_layout}(a), is 31\% smaller than the 6T SRAM layout depicted in Fig.~\ref{fig:gc_sram_layout}(b). The 2T OS-OS GCRAM layout, shown in Fig.~\ref{fig:gc_sram_layout}(b), is 89\% smaller than the 6T SRAM. Both 2T Si-Si GCRAM and 6T SRAM layouts are designed using logic design rules. It is worth noting that the GCRAM cell area can be further optimized by merging the connections of GND and dummy WLs with the power rail. The 2T OS-OS GCRAM is fabricated between tight-pitched metal layers, and thus we make the layout design meets the front-end-of-line (FEOL) design rules as defined in TSMC N40 PDK regarding the width, space, enclosure and extension. Moreover, the GCRAM device layers can be monolithically 3D stacked on top of the FEOL layers, taking no Si area budget at all to enable more area saving. Additionally, we resolved all DRC and LVS errors during the generation of GCRAM banks, with capacities ranging from 256 bits to 16 Kb.

\begin{figure}[ht]
    \centering
    \includegraphics[width=0.475\textwidth]{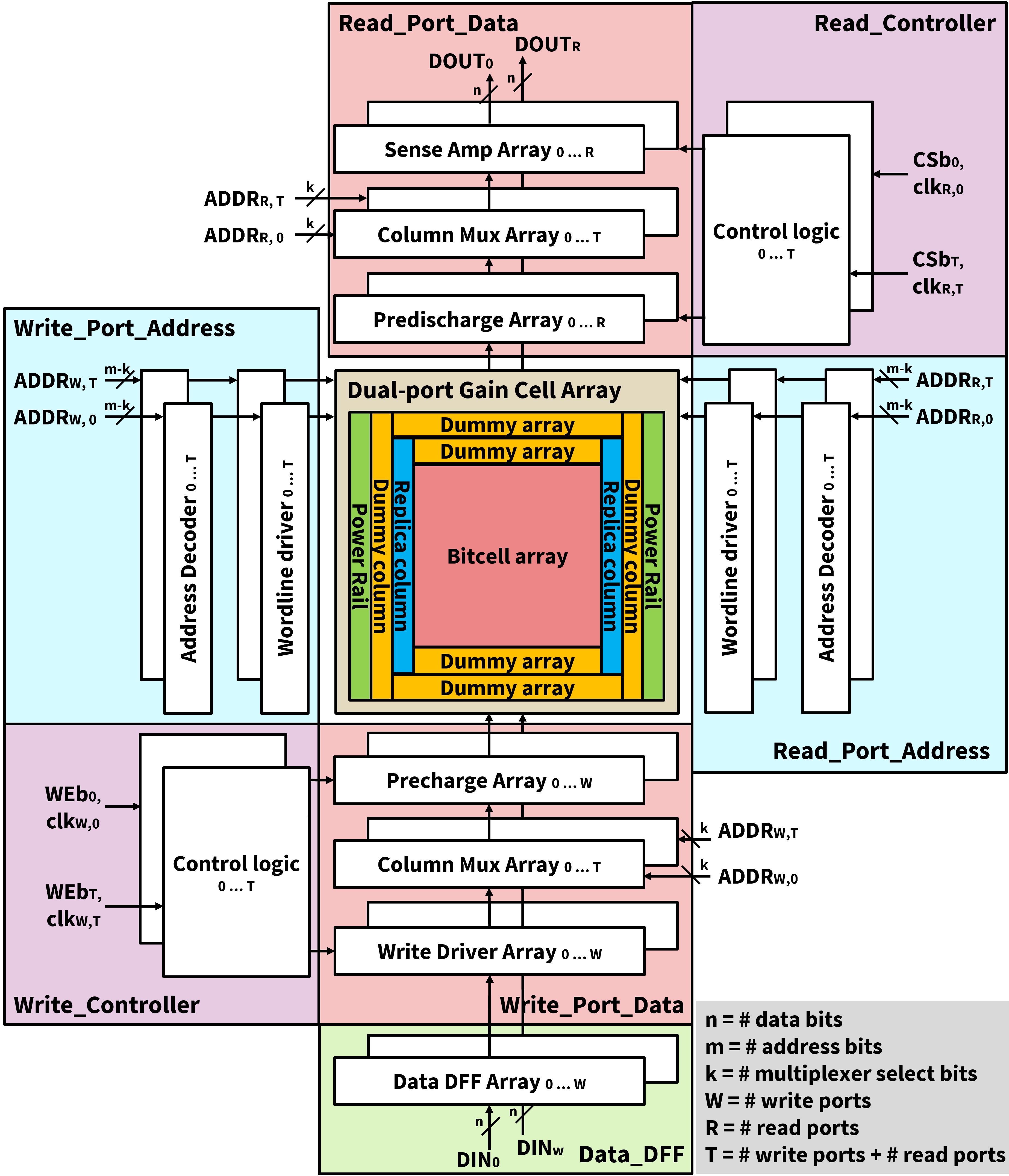} 
    \caption{The Architecture of a GCRAM bank. The OpenGCRAM compiler can automatically generate SPICE netlists and GDS layouts of all the modules and perform top-level integrations and simulations.}
    \label{fig:gc_arch}
    \vspace{-5pt}
\end{figure}

The GCRAM bank architecture is illustrated in Fig.~\ref{fig:gc_arch}. With separate read and write ports, the design includes a Write\_Port\_Address module located to the left of the bitcell array to drive the WWLs, and a Read\_Port\_Address module on the right to drive the RWLs. Similarly, a Write\_Port\_Data module at the bottom drives the WBLs, while a Read\_Port\_Data module at the top handles readout from the RBLs. the \url{Data\_DFF} latches the input data and holds it stable while it is written to the GCRAM bitcell. Additionally, the architecture includes two sets of control logic to generate the write and read enable signals independently.

Following the steps outlined in Fig.~\ref{fig:flow_chart}(b), we modified the relevant Python scripts to adjust the configurations of BLs and WLs and their connections to the peripheral circuits. Several changes were also made to the peripheral circuit design compared to OpenRAM. First, since the read and write operations in GCRAM are single-ended, the write driver and sense amplifier are connected to only one BL. The transistors and pins driving the $\mathrm{BL_{b}}$ in the write driver were removed. In the sense amplifier, the original $\mathrm{BL_{b}}$ connection was replaced with a connection to a reference voltage, which is generated by adding a reference generator \cite{Vref} module to the read control logic.

In the NMOS-NMOS Si-Si GCRAM configuration, the RWL is active-low, in which the falling edge of RWL signal will couple to storage node (SN) and cause SN voltage to degrade further in addition to WWL coupling degradation. An effective solution to this problem is to use complementary transistors for the read and write operations \cite{complementary_gc2, complementary_gc1, hybrid_gc}. In the NMOS-PMOS Si-Si GCRAM configuration, the RWL is active-high, in which the rising edge will boost the SN voltage and recover the voltage drop caused by WWL coupling. Unlike SRAM, where a precharge circuit maintains the RBL at a high voltage before RWL activation, Si-Si GCRAM relies on a predischarge circuit to ground the RBL before performing read operations. On the other hand, both OS transistors in OS-OS GCRAM are n-type, as the performance of p-type OS transistors is significantly worse than that of n-type transistors. Therefore, the OS-OS GCRAM bank still includes a precharge circuit for the read port.

To address this, we added a Python script to generate the NMOS circuits required for predischarging the RBL and modified the functions responsible for generating \url{Read\_Port\_Data} to connect the output pins of the predischarge array to the bitcell array. Additionally, unlike the precharge circuit, which is driven by an active-low $\mathrm{EN_{b}}$ signal, the predischarge circuit is controlled by an active-high EN signal. To accommodate this difference, we incorporated an inverter into the existing module responsible for generating the precharge $\mathrm{EN_{b}}$ signal within the read controller.

\begin{figure}[ht]
    \centering
    \includegraphics[width=0.475\textwidth]{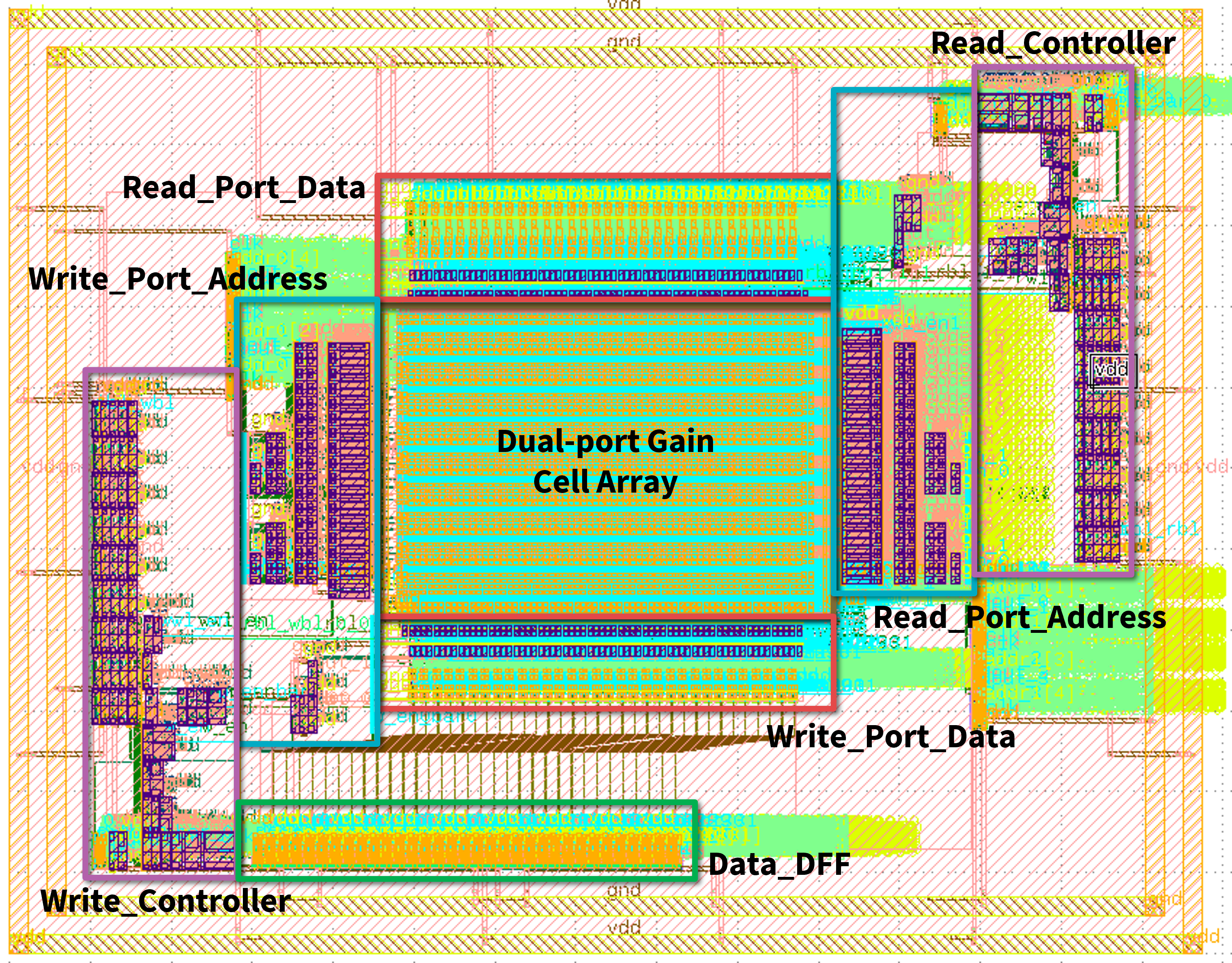} 
    \caption{The layout of a 32x32 GCRAM bank generated by OpenGCRAM. It consists a bitcell array, dual-port peripheral circuits and power rings and is ready for tapeout with TSMC.}
    \label{fig:layout}
\end{figure}
\subsection{Area Comparison}

\begin{figure}[ht]
    \centering
    \includegraphics[width=0.475\textwidth]{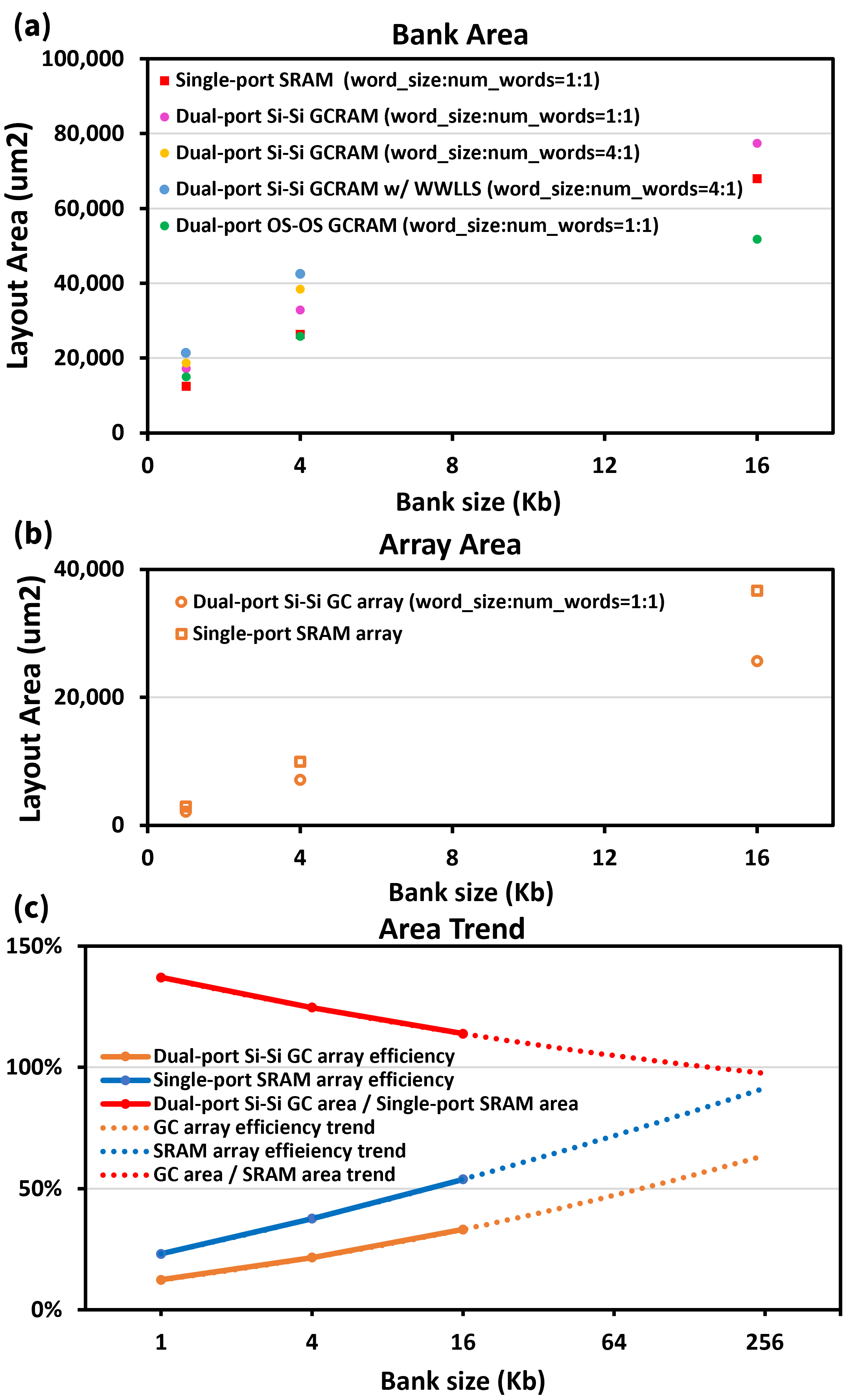} 
    \caption{The area comparisons between dual-port GCRAM and single-port SRAM. (a) Si-Si GCRAM initially has a larger area due to its dual-port peripheral circuits, whereas the OS-OS GCRAM banks are smaller since only the peripheral circuits consume silicon area. (b) The Si-Si GCRAM array is smaller than the SRAM array. (c) As the bank size increases, the area advantage of SRAM diminishes, as the amortization of peripheral circuitry becomes more significant, despite the comparison between dual-port and single-port configurations.}
    \label{fig:area}
    \vspace{-6.5pt}
\end{figure}

An example of GCRAM bank layout generated by the OpenGCRAM compiler is shown in Fig.~\ref{fig:layout}, featuring a 32x32 dual-port Si-Si GCRAM array and its peripheral circuits. A comparison between dual-port GCRAM banks and single-port 6T SRAM banks is presented in Fig.~\ref{fig:area}(a), covering bank sizes of 1 Kb, 4 Kb, and 16 Kb. As illustrated in Fig.~\ref{fig:gc_sram_layout}, a GCRAM bitcell is smaller than a 6T SRAM. 
However, the dual-port peripheral circuits in Si-Si GCRAM result in a larger overall bank size compared to a single-port 6T SRAM bank with the same word size (word\_size) and number of words (num\_words). The bitcell array area comparison is shown in Fig.~\ref{fig:area}(b), where the advantage of the Si-Si GCRAM array area becomes more pronounced as the bank size increases, owing to the smaller proportion of power rail area in the bitcell array. 
On the other hand, OS-OS GCRAM offers improved density, as the OS transistor can be fabricated in the back-end-of-line (BEOL) and monolithically 3D stacked on top of silicon circuits, enabling further reduction in bitcell area. OS-OS GCRAM banks occupy a smaller area than the SRAM banks, as only the peripheral circuits consumes silicon area, as shown in Fig.~\ref{fig:area}(a).
In Fig.~\ref{fig:area}(c), we plot the array efficiency of Si-Si GCRAM and 6T SRAM, along with the ratio of Si-Si GCRAM bank area to 6T SRAM bank area. Polynomial trend lines for the three curves are added and extrapolated to 64 Kb and 256 Kb. The results indicate that the GCRAM bank area eventually becomes smaller than that of SRAM with bank size larger than 256 Kb, highlighting the potential of GCRAM as a promising candidate for practical on-chip memory replacement.
Additionally, the dual-port feature of GCRAM allows simultaneous read and write operations, resulting in higher bandwidth. Note that, the area of a dual-port SRAM bank is approximately twice that of a single-port SRAM \cite{dual-port_sram} and remains larger than that of a dual-port Si-Si GCRAM for a fair comparison.

\subsection{Frequency, Bandwidth and Power Comparison}
\begin{figure}[ht]
    \centering
    \includegraphics[width=0.45\textwidth]{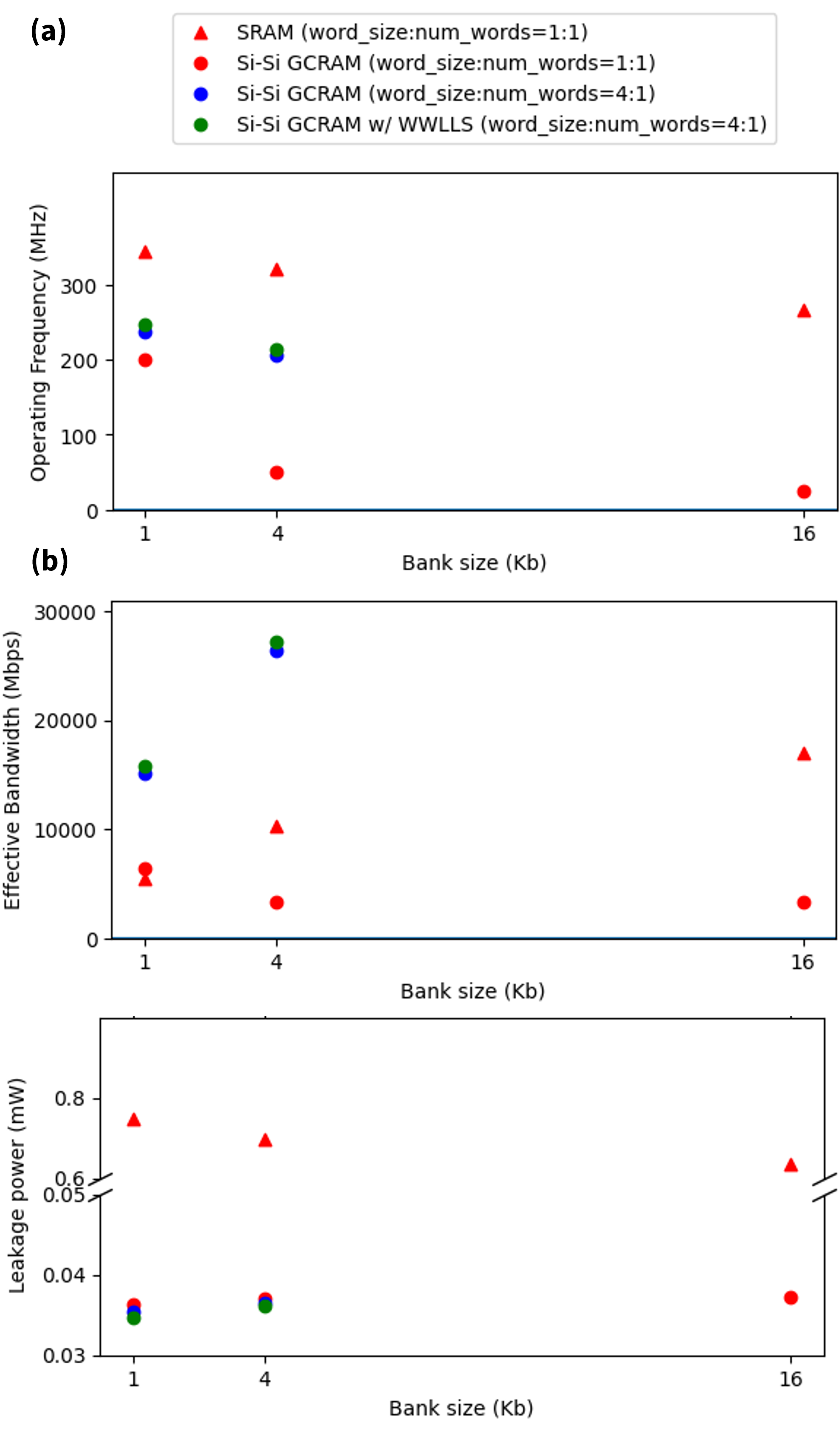} 
    \caption{(a) Operating frequency, (b) effective bandwidth and (c) leakage power comparison between Si-Si GCRAM and SRAM. With options to explore different design configurations, GCRAM can have comparable bandwidth and much lower leakage power compared with SRAM.}
    \label{fig:freq}
\end{figure}

OpenGCRAM supports both fast analytical delay and power estimations as well as precise HSPICE simulations. We evaluated the operating conditions of Si-Si GCRAM and SRAM, leveraging the compiler's ability to automatically generate stimuli, configure measurement setups, invoke the HSPICE simulator, and perform simulations on the generated SPICE netlist. Compared to GEMTOO, this SPICE simulation feature of OpenGCRAM provides more accurate performance evaluations for GCRAM. The operating frequency simulation results are summarized in Fig.~\ref{fig:freq}(a). Since the write speed of GCRAM and SRAM is higher than their read speed, the operating frequency is primarily constrained by the read operation. The operating frequency of Si-Si GCRAM is lower than that of SRAM.  This is likely due to the single-ended read operation of GCRAM. This issue can be addressed by optimizing the design of sense amplifiers and the read controller to better suit GCRAM applications. The sharp decrease in GCRAM frequency with word\_size:num\_words=1:1 observed between 1 Kb and 4 Kb is due to the additional delay chain stages in the read controller required to maintain correct timing conditions.
We also find out that GCRAM bank with the same bank size but different word\_size and num\_word parameters has different operating frequencies. When word\_size:num\_word=1:1, the compiler will put column mux circuits at the end of the BLs to force the organization of the bitcell array to be square. While when the word\_size:num\_word=4:1, the array will be naturally square and there is no need to add column mux circuits. Hence, the latter configuration results in better read speed, also shown in Fig.~\ref{fig:freq}(a). 

Additionally, the SN voltage only reaches $\mathrm{VDD - V_{T}}$ when writing data 1, resulting in a slower read speed. To boost the performance of GCRAM, we provide the option of adding a WWL level shifter (WWLLS) to the peripheral circuit design. By raising the WWL voltage to higher than VDD value, the SN voltage when writing data "1" increases, enabling higher speed in read operation. The results are shown as the green data points in Fig.~\ref{fig:freq}(a). However, adding another power supply to the circuit design results in an area penalty on the layout generation since another power ring needs to be added, as shown in Fig.~\ref{fig:area}(a).

We also compared the effective bandwidth of Si-Si GCRAM and SRAM, as shown in Fig.~\ref{fig:freq}(b). The effective read and write bandwidths of 6T SRAM are halved as the read and write operations share the same port and use the same frequencies in the HSPICE simulations. Overall, SRAM bandwidth is higher than GCRAM with the same configuration. However, GCRAM performance can be further optimized by adjusting the memory bank configurations and circuit designs.

The leakage powers of Si-Si GCRAM and SRAM, also simulated using netlists and stimuli automatically generated by OpenGCRAM compiler, are summarized and compared in Fig.~\ref{fig:freq}(c). Since there is no direct path from VDD to GND in the GCRAM bitcell, its leakage power is negligible and significantly lower than that of SRAM.

\hspace{2pt}
\subsection{GCRAM Retention Modulation}
\begin{figure}[ht]
    \centering
    \includegraphics[width=0.47\textwidth]{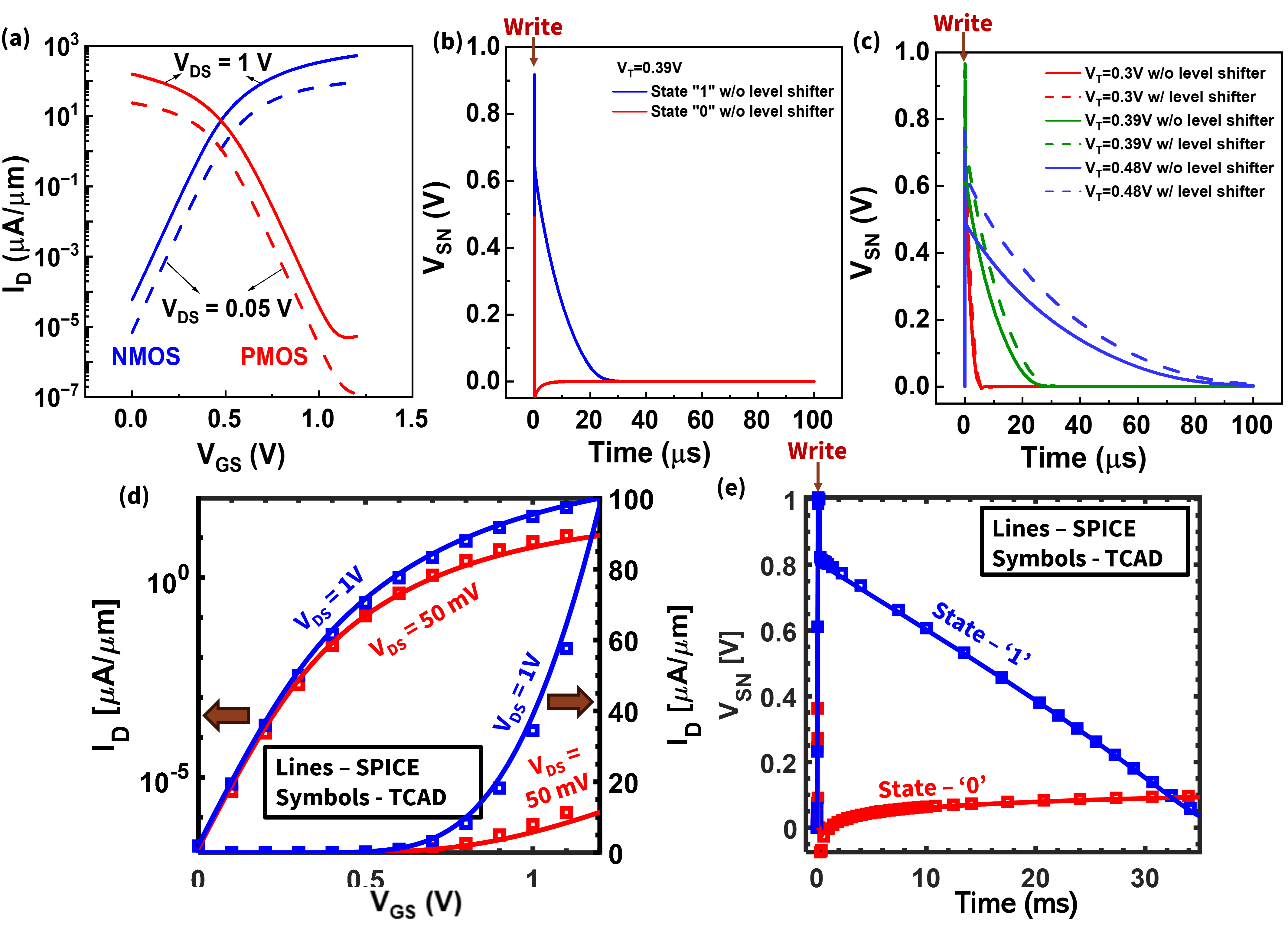} 
    \caption{(a) The $I_d-V_g$ properties of Si transistors. (b) Data written to Si-Si GCRAM can be retained for microseconds. (c) The retention of Si-Si GCRAM can be enhanced by adjusting the $V_T$ of the write transistor and adding a WWL level shifter. (d) The $I_d-V_g$ properties of OS transistor. (d) The OS-OS GCRAM exhibits much longer retention time because of its ultra-low leakage current.}
    \label{fig:retention}
\end{figure}
The retention time of GCRAM is primarily determined by the subthreshold channel leakage of the write transistor. Consequently, it can be modulated by adjusting the transistor design. Fig.~\ref{fig:retention}(a) illustrate the simulated $I_{d}-V_{g}$ properties of Si NMOS and PMOS transistors with TSMC N40 technology. Fig.~\ref{fig:retention}(b) presents the simulation results of post-write SN voltage, showing a microseconds-level retention time, primarily constrained by the decay of state "1". The retention time can be modulated as a function of the threshold voltage of the NMOS write transistor, as shown in Fig.~\ref{fig:retention}(c). As the subthreshold voltage increases, channel leakage decreases, allowing the SN charge to be retained for a longer duration at the cost of slower write and read operations. The simulation results also show that the retention can be enhanced with WWLLS included in the peripheral circuit. 

From a device perspective, the retention time of GCRAM is limited by stored charge leakage, which is constrained by the SN capacitance, the subthreshold leakage of the write transistor, and the gate dielectric leakage of the read transistor. To address this challenge, OS materials such as IGZO, ITO, and IWO have been extensively explored for the write transistor channel \cite{lu2022first, complementary_gc3,liu2023gain,ye2020double}. These materials exhibit extremely low off-state leakage currents (\textless1E-18 A/$\mathrm{\mu m}$), enabling long data retention and offering a much larger design space. We have conducted a TCAD simulation calibrated using experimental ultra-low leakage ITO transistor, and developed and calibrated a SPICE-compatible verilog-A model to fit the TCAD results. The $I_d-V_g$ plots of the OS transistor is shown in Fig.~\ref{fig:retention}(d). The simulation results shown that the retention time of OS-OS GCRAM, implemented with this ultra-low leakage ITO transistor, can be extended to the millisecond range, as depicted in Fig.~\ref{fig:retention}(e). Moreover, the OS-OS GCRAM retention can be enhanced to \textgreater10 s by increasing the $V_T$ through transistor/material engineering\cite{liu2023gain}. This tunable retention capability makes GCRAM versatile for various applications, such as activation caches and weight memory in AI inference. For instance, activation caches require microsecond-scale data lifetimes, while weight memory demands data lifetimes of hours \cite{yue2024wkvquant}.


\subsection{Design Space Exploration for AI workloads}
\begin{figure}[ht]
    \centering
    \includegraphics[width=0.4\textwidth]{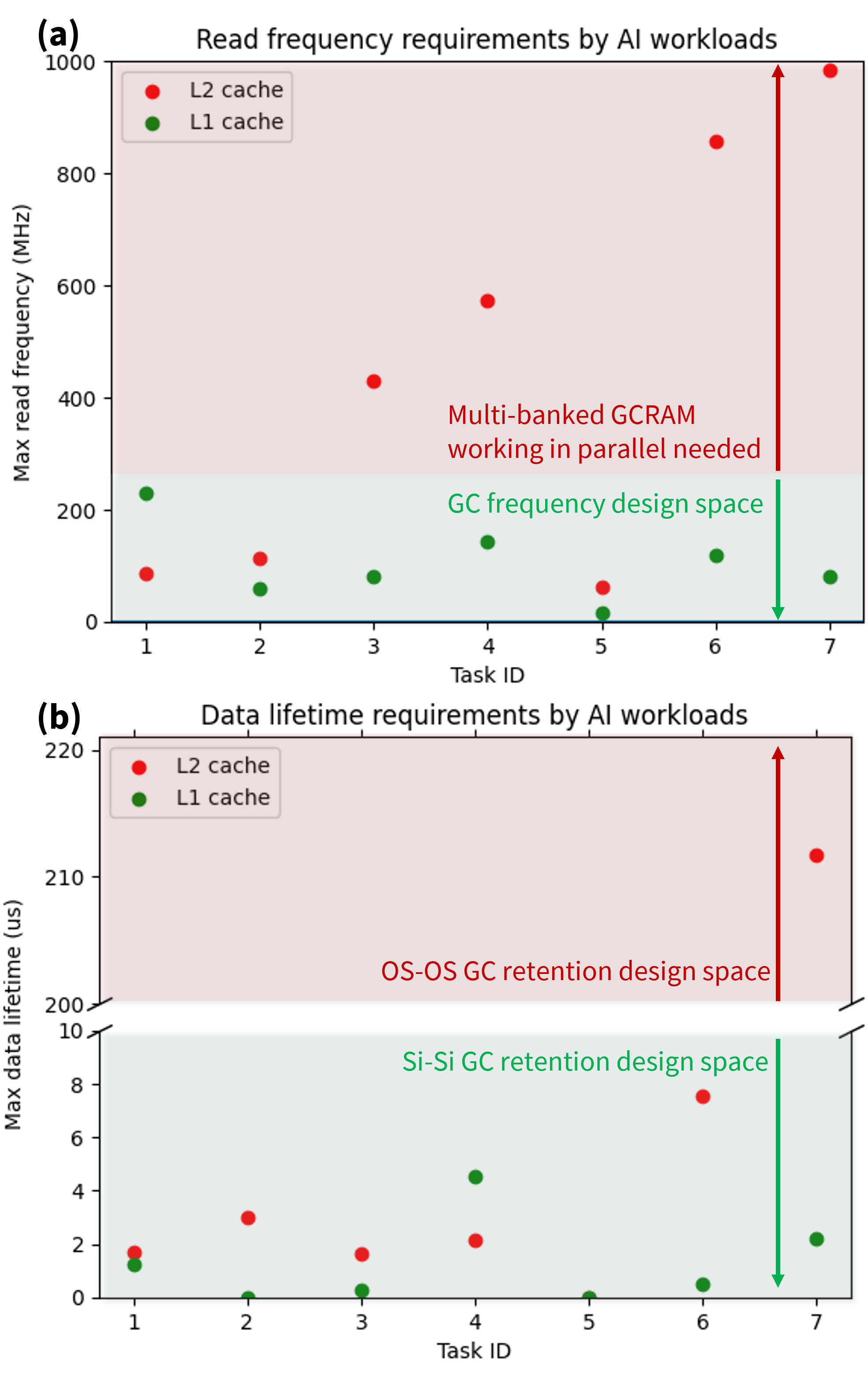} 
    \caption{The demands on L1 and L2 caches in terms of (a) read frequency and (b) data lifetime for implementing various AI workloads. OpenGCRAM enables the generation of optimal GCRAM bank designs tailored to specific performance metrics through design-space exploration.}
    \label{fig:l1_l2_cache}
\end{figure}

\begin{figure}[ht]
    \centering
    \includegraphics[width=0.45\textwidth]{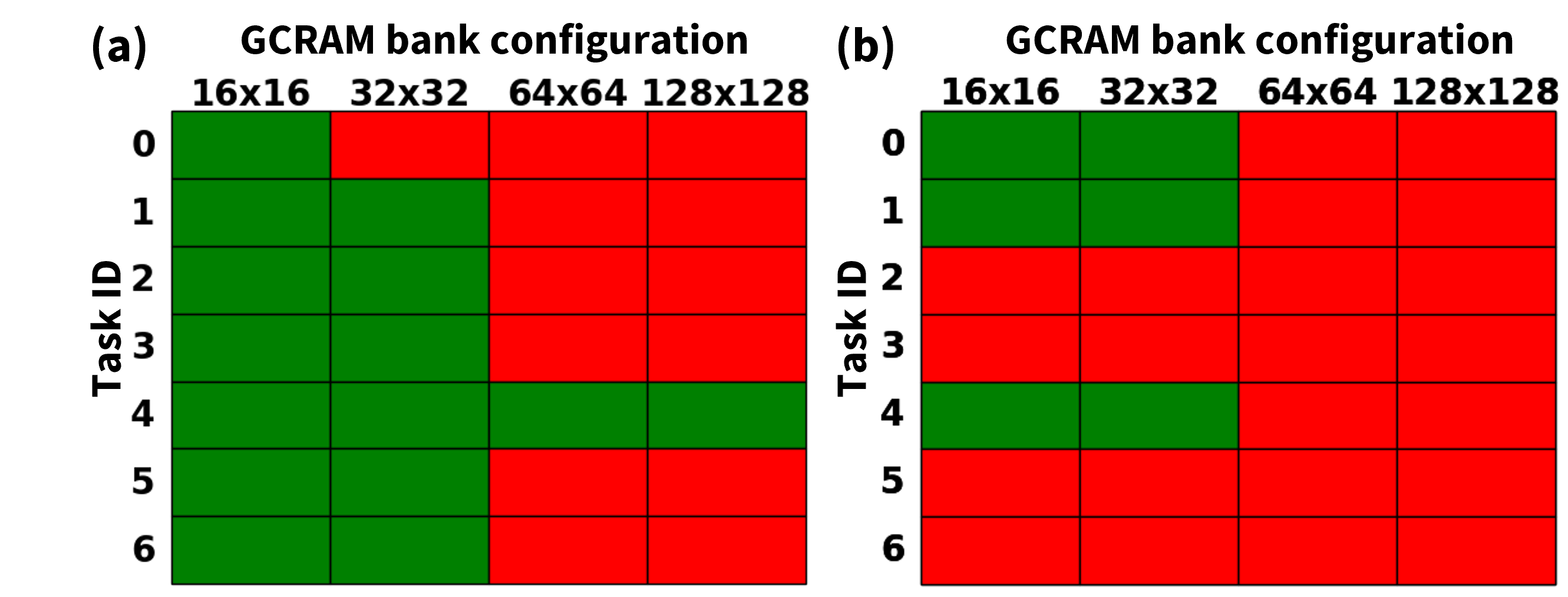} 
    \caption{Design choices for implementing (a) L1 and (b) L2 caches with single-banked GCRAM for various AI workloads. The plot demonstrates how OpenGCRAM facilitates design-space exploration to optimize GCRAM bank parameters for specific applications.}
    \label{fig:rshmoo}
\end{figure}

\begin{table}[t!]
    \centering
    \caption{AI workloads evaluated by the GainSight profiler}
    \label{tab:ai_workloads_id}
    \resizebox{\columnwidth}{!}{%
    \begin{tabular}{p{0.3cm}lll}
    \toprule
    \multicolumn{1}{l}{\textbf{\begin{tabular}[c]{@{}l@{}}Task \\ ID\end{tabular}}} &
    \multicolumn{1}{l}{\textbf{Task Name}} & \multicolumn{1}{l}{\textbf{Test Suite}} & \multicolumn{1}{l}{\textbf{Description}} \\ \midrule
    \texttt{1} & 2dconvolution & PolyBench~\cite{grauer2012auto,pouchet2013polyhedral} & 2D Convolution \\ \midrule
    \texttt{2} & 3dconvolution & PolyBench & 3D Convolution \\ \midrule
    \texttt{3} & llama-3.2-1b & ML Inference~\cite{reddi2020mlperf} & \begin{tabular}[c]{@{}l@{}}Meta's text-based LLM with 1 \\ billion parameters~\cite{touvron_llama_2023}\end{tabular} \\ \midrule
    \texttt{4} & llama-3.2-11b-vision & ML Inference & \begin{tabular}[c]{@{}l@{}l@{}}Meta's LLM with integrated vision \\ adapter for image recognition, \\ total of 11 billion parameters~\cite{touvron_llama_2023}\end{tabular} \\ \midrule
    \texttt{5} & resnet-18 & ML Inference & \begin{tabular}[c]{@{}l@{}}CNN for image recognition with \\ 18 layers~\cite{he_deep_2015} \end{tabular}\\ \midrule
    \texttt{6} & bert-uncased-110m & ML Inference & \begin{tabular}[c]{@{}l@{}l@{}}"Bidirectional Encoder Representation \\ for Transformers"~\cite{devlin2019bert}, text-based \\ LLM with 110 million parameters\end{tabular} \\ \midrule
    \texttt{7} & stable-diffusion-3.5b & ML Inference & \begin{tabular}[c]{@{}l@{}}Text-to-image transformer model with \\ 3.5 billion parameters~\cite{esser2024scaling} \end{tabular} \\ \bottomrule
    \end{tabular}%
    }
\end{table}
The OpenGCRAM compiler enables fast and accurate GCRAM bank generation and performance simulation, making it easy to perform design-space exploration to meet the requirements of various workloads. Using the GainSight profiling framework \cite{gainsight}, we have extracted the performance demands for L1 and L2 caches across different AI workloads. The list of evaluated AI tasks is provided in Table~\ref{tab:ai_workloads_id}. The maximum read frequencies and data lifetime required for the L1 cache and L2 cache are summarized in Fig.~\ref{fig:l1_l2_cache}, with the profiling evaluated on NVIDIA H100 and scaled for NVIDIA GeForce GT 520M. 
In Fig.~\ref{fig:rshmoo}, we present Shmoo plots that illustrate the design choices of the GCRAM bank for different tasks with different configurations. The X-axis represents the bank size (word\_size multiplied by num\_words), spanning from 16x16 to 128x128, while the Y-axis shows the task ID in Table~\ref{tab:ai_workloads_id}. The data points in the plot represent whether the bank can successfully work or not.

The plot reveals that a GCRAM bank no larger than 1 Kb works for most L1 cache applications and several of the L2 cache applications, since they can operate with higher frequency. When a specific task can be accomplished with multiple configurations, larger bank size is better since it can offer higher bandwidth and higher effective memory density. In addition, Si-Si GCRAM banks can retain the data long enough to meet lifetime requirements except for the L2 cache for the stable diffusion task. Counterintuitively, most of L2 tasks require a much higher read frequency than the L1 cache. This is because the L2 cache is shared by all SIMD cores in the GPU, while each core has its own dedicated L1 cache. Hence, the L2 cache needs to handle much more read and write requests from multiple cores. 
Analogous to how NVIDIA GPUs organize the L2 SRAM cache, we can employ a multi-banked GCRAM design to accommodate for multiple parallel read and write requests. This information can be used for fine-tuning the parameters used in generating GCRAM bank to optimize for specific applications, indicating the ability to perform a design-space exploration in a fast, accurate and optimized approach.


\section{Future Work}
The methodology for extending the compiler to support new PDKs and memory technologies has paved the way for expanding OpenGCRAM’s functionality to accommodate various types of GCRAM. In addition to the Si-Si GCRAM, hybrid OS-Si GCs (featuring OS write transistors and Si read transistors) \cite{hybrid_gc} and OS-OS GCRAM are additional options.  

    

OpenGCRAM is designed to flexibly incorporate these GCRAM variants and choose from them based on different applications. We have incorporated the netlist and layout generation of OS-OS GCRAM bank into OpenGCRAM. The HSPICE simulation will also be implemented to evaluate the operating frequency, bandiwdth and power of OS-OS GCRAM. Hybrid Si-OS GCRAM will also be added as its performance can cover the design space between Si-Si and OS-OS GCRAM by offering moderate retention and frequencies. Additional custom circuit modules (such as the current mode sense amplifier \cite{csa}), will also be incorporated to enhance the performance of GCRAM (such as read speed). Given the performance requirements, OpenGCRAM can explore various GCRAM variants and peripheral circuit designs to determine the optimal configuration. The design space of GCRAM will be further expanded by enabling multibank GCRAM generation. Additionally, we plan to implement area-delay-power co-optimization within OpenGCRAM, leveraging machine learning algorithms (e.g., gradient descent) to optimize configurations for specific application targets. Note that this methodology is not limited to TSMC 40 nm technology and GCRAM. It can be extended to support more advanced technology nodes, such as TSMC 16 nm, and other memory technologies, including 1T1C DRAM and RRAM.

\section{Conclusion}
In this paper, we introduced OpenGCRAM, an open-source GCRAM compiler, and presented a modular methodology for porting the OpenRAM compiler to new PDKs and memory technologies. This approach enables the generation of GCRAM circuit designs and tapeout-ready layouts, as demonstrated with TSMC 40nm technology. Additionally, OpenGCRAM supports HSPICE simulations to evaluate the operating conditions of GCRAM banks. By providing fast and accurate automatic GCRAM bank generation and simulation, OpenGCRAM facilitates design-space exploration, enabling the selection of optimal memory configurations tailored to specific AI workload performance metrics. Looking ahead, we envision extending this compiler to incorporate optimization techniques based on machine learning algorithms to dynamically select the best configuration according to performance targets. Furthermore, our methodology for implementing OpenGCRAM can be applied to support more advanced technology nodes, additional GCRAM variants, and other memory technologies.

\newpage
\bibliographystyle{IEEEtran}
\bibliography{references}
\vspace{12pt}
\end{document}